\documentclass[11pt,twoside]{article}
\usepackage{asp2010}

\resetcounters
\usepackage{graphics}
\usepackage{graphicx}

\newcommand{\cmt}{\,cm$^{-3}$}
\newcommand{\cmd}{\,cm$^{-2}$}
\newcommand{\lo}{\,$L_{\sun}$}
\newcommand{\ro}{\,$R_{\sun}$}

\begin{document}

\title{Multiwavelength modeling the SED of very slow novae 
       PU~Vul and V723~Cas}
\author{Augustin Skopal
\affil{Astronomical Institute, Slovak Academy of Sciences, 
       Tatransk\'a Lomnica}}

\begin{abstract}
Evolution in the spectrum of very slow novae PU~Vul and V723~Cas 
during their transition from the optical maximum to the nebular 
phase is investigated using the method of disentangling the 
composite UV/optical spectra. 
Model SEDs suggested that a transient decrease in the WD 
luminosity, during the decline from the maximum, was caused by 
a negative beaming effect, when a neutral disk around the WD 
was formed. When the disk disappeared, the luminosity increased 
again to values from the beginning of the outburst (in the case 
of V723~Cas, at/above the Eddington limit). This suggests the 
presence of a mechanism maintaining a high energy output for 
a much longer time than it is predicted by the current theories. 
Similarity of LCs, but enormous difference of the separation    
between the components of PU~Vul and V723~Cas binaries suggest     
that the mechanism is basically powered by the accretor. 
\end{abstract}

\section{Introduction}

Novae represent an observable result of thermonuclear outburst 
ignited on the surfaces of white dwarfs (WDs). The burning 
material is accreted from a companion star in binary. 
In the case of Classical Novae (CNe) the donor star is 
usually a main-sequence red dwarf transferring matter onto 
its WD companion via the Roche-lobe overflow. Orbital periods 
of CN binaries are in order of hours. 
In the case of Symbiotic Novae (SNe) the thermonuclear runaway 
phenomena are occurring in binary systems that consist of 
a WD and a red giant. Orbital periods are in order of years, 
and the WDs are accreting from the giant's wind. 
Evolution of the outburst is usually best documented by the 
optical photometry. It has demonstrated very different profiles 
of the novae light curves (LCs). To quantify the rates of decline 
from maximum, the speed classes (from very fast to very slow) 
were introduced \citep[e.g.][and references therein]{bode+08}. 

In this contribution I demonstrate nova evolution, during the 
transition from the optical maximum to the nebular phase, for 
two very different types of novae (CN and SN), but with similar LCs. 
For this purpose I selected the SN PU~Vul and CN V723~Cas, whose 
LCs were classified as very slow. Diversity of types, but basic 
similarity of their LCs can aid us in better understanding 
the origin of common properties of their SEDs. 

\section{Multiwavelength modeling the SED}

According to observational properties of novae, their SED in 
the UV/optical continuum, $F(\lambda)$, can be expressed as 
a superposition of three basic radiative components, i.e., 
%
\begin{equation}
  F(\lambda) =  F_{\rm WD}(\lambda) + F_{\rm N}(\lambda) +
                F_{\rm G}(\lambda),
\label{eq:1}  
\end{equation}
where $F_{\rm WD}(\lambda)$ is the flux produced by the WD's 
pseudophotosphere, $F_{\rm N}(\lambda)$ is the nebular 
component of radiation from thermal plasma and 
$F_{\rm G}(\lambda)$ represents the contribution from the giant 
in SNe. 
For low temperatures, $T^{\rm eff}_{\rm WD} \sim 5000-15000$\,K, 
an atmospheric model, 
$\mathcal{F}_{\lambda}(T^{\rm eff}_{\rm WD}$), is usually 
required to fit the optical continuum. For higher temperatures, 
blackbody radiation can be used to match the UV continuum. 
From absorption events, acting in the ejecta, we quantify only 
the attenuation caused by the Rayleigh scattering of the far-UV 
photons on the neutral atoms of hydrogen. 
The nebular radiation in the continuum can be approximated by 
processes of recombination and thermal bremsstrahlung in the 
hydrogen plasma for {\em Case}~B. 
Finally, radiation from the giant is represented by an 
appropriate synthetic spectrum, 
$\mathcal{F}_{\lambda}(T^{\rm eff}_{\rm G})$. 
Then Eq.~(\ref{eq:1}) can be written in a form, 
%
\begin{equation}
 F(\lambda) =   
      \theta_{\rm WD}^2 \mathcal{F}_{\lambda}(T^{\rm eff}_{\rm WD})\,
      e^{-\sigma_{\rm Ray}(\lambda)\,N_{\rm H}} +    
      k_{\rm N} \varepsilon_{\lambda}({\rm H},T_{\rm e}) +
      \theta_{\rm G}^2 \mathcal{F}_{\lambda}(T^{\rm eff}_{\rm G}),
\label{eq:2}
\end{equation}
%
where scalings
$\theta_{\rm WD} = R_{\rm WD}/d$ 
and 
$\theta_{\rm G} = R_{\rm G}/d$ 
represent angular radii of the WD pseudophotosphere and the giant, 
respectively. The factor $k_{\rm N}$ scales the volume 
emission coefficient $\varepsilon_{\lambda}(T_{\rm e})$ 
of the nebular continuum to observations. Constant $T_{\rm e}$ 
throughout the nebula is assumed. $N_{\rm H}$ is the hydrogen 
column density and $\sigma_{\rm Ray}(\lambda)$ is the Rayleigh 
scattering cross-section. 

In the SED-fitting analysis, model variables 
($\theta_{\rm WD}$, $\theta_{\rm G}$, 
$T^{\rm eff}_{\rm WD}$, $T^{\rm eff}_{\rm G}$, 
$N_{\rm H}$, $k_{\rm N}$ and $T_{\rm e}$) 
are given by the solution of Eq.~(\ref{eq:2}), which corresponds 
to a minimum of the reduced $\chi^{2}$ function. Atmospheric 
models for $T^{\rm eff}_{\rm WD}$ = 5000--15000\,K were taken 
from \cite{m+z02} and those for giants from \cite{fluks+94}. 
More details can be found in \cite{sk05}. 

\section{Results}

\subsection{Symbiotic nova PU~Vul}

PU~Vul is a well known very slow eclipsing symbiotic nova 
that exploded at the end of 1977. 
Its LC shows a $\sim$9-year-lasting flat maximum with some 
erratic brightenings, followed by a slow decline from the 
beginning of 1988. 
It is a binary system comprising an M giant and a WD accreting 
from the giant's wind on a 13.4-yr orbit 
\citep[e.g.][]{belyakina+82,kolotilov+95,shugarov+12}. 
Recently, \cite{kato+12} performed a model of the $V$-LC 
consisting of the emission from the outbursting WD, its M-giant 
companion and the nebulae. Using the model LC, they specified 
new values of reddening $E_{\rm B-V} \sim 0.3$\,mag and 
distance $d\sim 4.7$\,kpc. 

During the flat maximum, an atmospheric model for 
$T^{\rm eff}_{\rm WD}\sim 7000$\,K matches well the observed 
SED. It corresponds to $L_{\rm WD}\sim 15800$\lo\ and 
$R_{\rm WD}\sim 86$\ro. 
During the decline phase, the two-temperature type of the 
UV/optical spectrum developed. On 1988 October 10, the measured 
luminosity decreased to $L_{\rm WD} \sim 10700$\lo\ 
($T^{\rm BB}_{\rm WD} \sim 20000$\,K, $R_{\rm WD} \sim 8.6$\ro), 
a strong nebular radiation with the emission measure 
$EM \sim 9.6 \times 10^{60}$\cmt\ was detected, and the far-UV 
continuum was attenuated by the Rayleigh scattering on 
$\sim 3.9\times 10^{22}$\cmd\ atoms of hydrogen. 
During quiescent phase, when the signatures of the Rayleigh 
scattering disappeared, the strong nebula 
($EM \sim 9.6 \times 10^{60}$\cmt) and the steep slope of the 
far-UV continuum required 
$T^{\rm BB}_{\rm WD} > 79000$\,K, $R_{\rm WD} < 0.64$\ro, 
resulting in an increase of $L_{\rm WD}$ to $> 14300$\lo. 
%
%
\begin{figure*}[!t]
\centering    
\begin{center}
%
\resizebox{\hsize}{!}{\includegraphics[angle=-90]{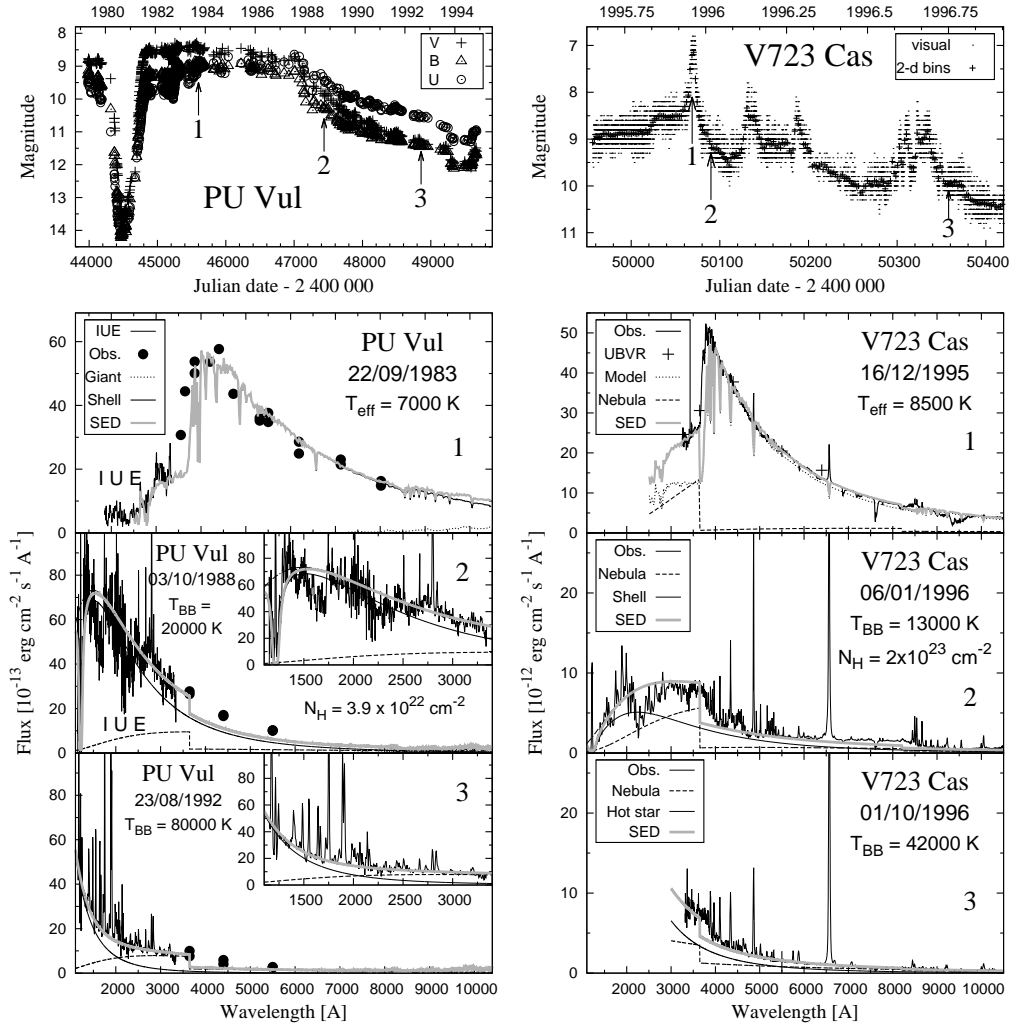}}
\caption[]{
Top panels show the LCs of PU~Vul and V723~Cas covering the 
maximum and following decline. The lower panels display the 
observed and model SED at dates marked by arrows in LCs. 
The \textsl{HPOL} spectrophotometry (320--1050\,nm) of V723~Cas 
and \textsl{IUE} spectra were obtained from archives by MAST. 
          }
\end{center}
\end{figure*} 

\subsection{Classical nova V723~Cas}

V723~Cas was discovered on 1995 August 24. The visual LC 
of V723~Cas was classified as very slow showing numerous 
$1-2$\,mag flares on the time-scale of days/weeks 
\citep[e.g.][]{munari+96,chochol+97}. The authors also noted 
its similarity to the LC of the SN PU~Vul. The orbital period 
of the V723~Cas binary was determined by \cite{chochol+00} 
to 16.6 h. 
The very slow evolution was also pointed by \cite{ness+08}, 
who found that V723~Cas was active X-ray source more than 12 
years after outburst. Another signature of such behaviour 
was revealed by \cite{schaefer+10}, who found that the 
post-eruption LC ($> 2003$) is brighter by $\sim 3$\,mag 
than before the eruption. 

The model SEDs performed during the same period of the 
LC-evolution are very similar to those of PU~Vul (Fig.~1). 
During the optical maximum (1995 December 16) the WD's 
pseudophotosphere was as large as 
$R_{\rm WD} \sim 95$\ro\ 
and radiated at 
$T^{\rm eff}_{\rm WD} \sim 8500$\,K, 
which corresponds to a super-Eddington luminosity 
$L_{\rm WD} \sim 42000$\lo\ ($d=3$\,kpc, 
$E_{\rm B-V} = 0.5$\,mag). In addition, a large amount of 
the nebular radiation ($EM \sim 2.2 \times 10^{61}$\cmt) 
was recognized by the model. 
At the decline, the two-temperature type of the UV/optical 
spectrum developed as in the case of PU~Vul. On 1996 January 6, 
the measured luminosity decreased to $L_{\rm WD} \sim 8700$\lo\ 
($T^{\rm BB}_{\rm WD} \sim 13000$\,K, $R_{\rm WD} \sim 14$\ro), 
a very strong nebula with $EM \sim 2.0 \times 10^{61}$\cmt\ 
was indicated, and the far-UV continuum was attenuated with 
the Rayleigh scattering ($N_{\rm H} \sim 2\times 10^{23}$\cmd). 
After the 1996 Aug./Sept. flare, the profile of the 320--1050\,nm 
spectrum required significantly hotter stellar source of radiation. 
The strong nebular emission with $EM \sim 2.6 \times 10^{61}$\cmt\ 
required $T^{\rm BB}_{\rm WD} > 42000$\,K and $R_{\rm WD} < 4$\ro, 
which means that the luminosity ($> 44000$\lo) increased again 
to/above the Eddington limit. The models are shown in Fig.~1.

\section{Concluding remarks}

During the optical maxima, the SED of the outbursting WDs was 
comparable with a 6--9\,kK model atmosphere scaled to the 
luminosity of $\sim$16000 and $\sim$42000\lo\ for PU~Vul and 
V723~Cas, respectively. 

During the decline from maxima, the SED shifted to higher 
energies having signatures of the neutral disk-like material 
between the observer and the burning WD. The two-temperature 
UV spectrum developed. The measured luminosity decreased, 
because of the undetectable fraction of the WD's radiation 
emitted to/around the pole directions 
\citep[a negative beaming effect, see Sect.~5.3.6 of][]{sk05}. 

When the neutral material disappeared, the luminosity increased 
to that from the beginning of outbursts. This suggests the 
presence of a mechanism maintaining the high energy output for
a much longer time than it is predicted by the current theories. 
Similarity of LCs, but very different separations of the binary 
components in PU~Vul and V723~Cas suggest that the mechanism 
is basically powered by the accretor. 

\acknowledgements 
This research was supported by a grant of the Slovak Academy 
of Sciences, VEGA No. 2/0002/13.

%
{}
\end{document}